\documentclass[%
 reprint,
superscriptaddress,
 amsmath,amssymb,
 aps,
floatfix,
]{revtex4-1}
\usepackage{graphicx}
\usepackage{dcolumn}
\usepackage{bm}
\usepackage{color}

\begin{document}

\preprint{APS/123-QED}

\title{\color{blue} Microscopic Theory of Onset of De-Caging and Bond Breaking Activated Dynamics in Ultra-Dense Fluids with Strong Short Range Attractions}
\author{Ashesh Ghosh}
\affiliation{Department of Chemistry}
\affiliation{Materials Research Laboratory}
\author{Kenneth S. Schweizer}
\email[]{\color{blue}kschweiz@illinois.edu}
\affiliation{Department of Chemistry}
\affiliation{Materials Research Laboratory}
\affiliation{Department of Material Science \& Engineering}
\affiliation{Department of Chemical \& Biomolecular Engineering \\ University of Illinois at Urbana-Champaign, Illinois 61801, USA}

\begin{abstract}
We theoretically study thermally activated elementary dynamical processes that precede full structural relaxation in ultra-dense particle liquids interacting via strong short range attractive forces. Our approach is based on a microscopic theory formulated at the particle trajectory level built on the dynamic free energy concept and an explicit treatment of how attractions control physical bonding. Mean time scales for bond breaking, the early stage of cage escape, and a fixed non-Fickian displacement are analyzed in the repulsive glass, bonded repulsive (attractive) glass, fluid, and dense gel regimes. The theory predicts a strong length-scale-dependent growth of these time scales with attractive force strength at fixed packing fraction, a much weaker slowing down with density at fixed attraction strength, and a strong decoupling of the shorter bond breaking time with the other two time scales that are controlled mainly by perturbed steric caging. All results are in good accord with simulations, and additional testable predictions are made. The classic statistical mechanical projection approximation of replacing all bare attractive and repulsive forces with a single effective force determined by pair structure incurs major errors for describing processes associated with thermally activated escape from transiently localized states.
\end{abstract}
\maketitle
How strong short range attractive forces control length scale dependent slow activated dynamics and kinetic arrest of soft matter systems is a problem of broad fundamental interest which continues to challenge theoretical understanding. The central issue is the non additive interplay of physical bond formation and steric caging in high packing fraction fluids and colloidal suspensions in determining structural relaxation, intermediate time non-Fickian motion, diffusion, and nonlinear rheology \cite{poon2004,Poon_2002,Pham104,Eckert2002,Royall2017,ZP2009,poon_europhys_2006,Puertas_2002,Zaccarelli_2008}. 
\par
Seminal studies based on microscopic ideal mode coupling theory (MCT) predicted a subtle structure-induced competition between the dynamical consequences of repulsive force caging and physical bonding, and led to the concepts of attractive (bonded) glasses and re-entrant glass melting \cite{Pham104,Puertas_2002,Zaccarelli_2008,gotze_2012}. However, MCT does not account for ergodicity restoring activated processes resulting in bonds and cages that persist forever which leads to some qualitative disagreements with simulation and experiment\cite{ZP2009,gotze_2012,Hunter_2012,Weeks627,Kob}. The non-linear Langevin Equation (NLE) theory\cite{Saltzman2003,kss2005}, and its more recent Elastically Collective NLE (ECNLE) extension\cite{mirigian2013} explicitly address activated motion by avoiding closure approximations for ensemble-averaged time correlation functions in favor of a stochastic trajectory level description. The central quantity is the “dynamic free energy”\cite{kss2005}, and relaxation via thermal noise driven barrier hopping has been shown to be well captured in repulsion-dominated colloidal \cite{mirigian2013,Steve12014}, molecular \cite{mirigian2013,Steve22014}, and polymeric \cite{mirigian2015,shijie2016} glass forming liquids. Dynamic constraints are quantified based on the “projection approximation” \cite{gotze_2012} that replaces all Newtonian forces with an effective force determined by the equilibrium pair correlation function. 
\par
Very recently we established that the projection approximation for dense strongly attractive particle fluids in the NLE or ECNLE theoretical framework results in qualitatively incorrect predictions for long time activated relaxation at the very high packing fractions characteristic of the attractive glass and dense gel regimes\cite{agkss2019}. A new approach (projectionless dynamics theory (PDT)) \cite{agkss2019,dellkss2015} for quantifying effective forces was formulated which explicitly treats how short range attractive forces create physical bonds and interfere with caging. Detailed comparisons with experiments and simulations suggest it properly captures re-entrant dynamic arrest, strong non-monotonic variation of the activated structural relaxation time and long time diffusion constant with attraction strength, and kinetic criteria for glass and gel formation based on a practical observation time\cite{agkss2019}. The physical picture is conceptually distinct from phenomenological attempts \cite{Puertas_2002,Zaccarelli_2008,amokrane2019,amokrane2010} to use ideal MCT to rationalize slow activated dynamics of dense attractive colloids. 
\par
To date, the ECNLE-PDT approach has only considered the slowest structural relaxation process associated with long time barrier crossing\cite{agkss2019,dellkss2015}. Since the theory is formulated in terms a spatially-resolved dynamic free energy\cite{Saltzman2003,kss2005}, earlier time motions on smaller “in cage” length scales can be naturally addressed. This Rapid Communication reports the first analysis of this aspect, which is of interest in its own right and provides a deeper test of both the concept of a dynamic free energy and direct treatment of attractive forces. Our work also serves as a microscopic basis for understanding the large scale simulations (over 8 decades in time) of Zaccarelli and Poon (ZP)\cite{ZP2009} where repulsive glass (RG), bonded repulsive glass (BRG) or attractive glass (AG), and dense gel (DG) regimes of slow dynamics exist. The absolute and relative times scales for the onset of de-caging and bond breaking, and how they evolve with density and attractive force strength and range, are analyzed. These phenomena occur well beyond the ideal MCT transition in a regime of strongly activated, highly non-Gaussian motion.
\par
To frame our work, we first summarize the ZP simulations and then further examine them from our theoretical perspective. These workers considered a binary mixture of hard spheres of different diameters that interact via a short range (3\% mean particle diameter) square well attraction of strength $\beta\epsilon=0-5k_BT$ at very high mixture packing fractions of $\phi=0.57-0.63$. The inset of Fig. 1b shows the specific state points studied, along with a idea MCT kinetic arrest map that has been empirically shifted both horizontally and vertically to qualitatively align with simulation deductions\cite{Sciortino2003}. The attraction strengths span the range from well below to well above the re-entrant “nose” feature of the MCT boundary associated with non-monotonic or re-entrant dynamics\cite{Pham104,ZP2009,Puertas_2002,Sciortino2003}. Figure 1a shows selected single particle mean square displacements (MSD) \cite{ZP2009} which are mostly sub-diffusive and involve displacements well below a particle diameter. The densest RG state 0 is a hard sphere (HS) glass where cages are stable (flat MSD) on the simulation time scale. BRG states I (not shown) and II exhibit a ‘temporal trapping’ on the attraction range length scale due to bond formation, followed by a slow increase towards the hard sphere fluid long-lived plateau. States III-V are highly attractive but at lower (still high in an absolute sense) packing fractions (DG) and show similar, but slightly weaker, short time dynamic localization. However, the latter states do not exhibit a MSD that saturates after the bond-breaking event. Weaker attractive states VI, VII (plotted) and VIII, deemed “fluids” (F) in ref. \cite{ZP2009}, show an inflection point in the MSD before crossing over to Fickian diffusion. The form of the MSD appears to continuously evolve with attraction strength and packing fraction, with the RG, BRG, DG and F states all exhibiting intermittent dynamics.
\par
\begin{figure}
    \centering
    \includegraphics[scale=0.40]{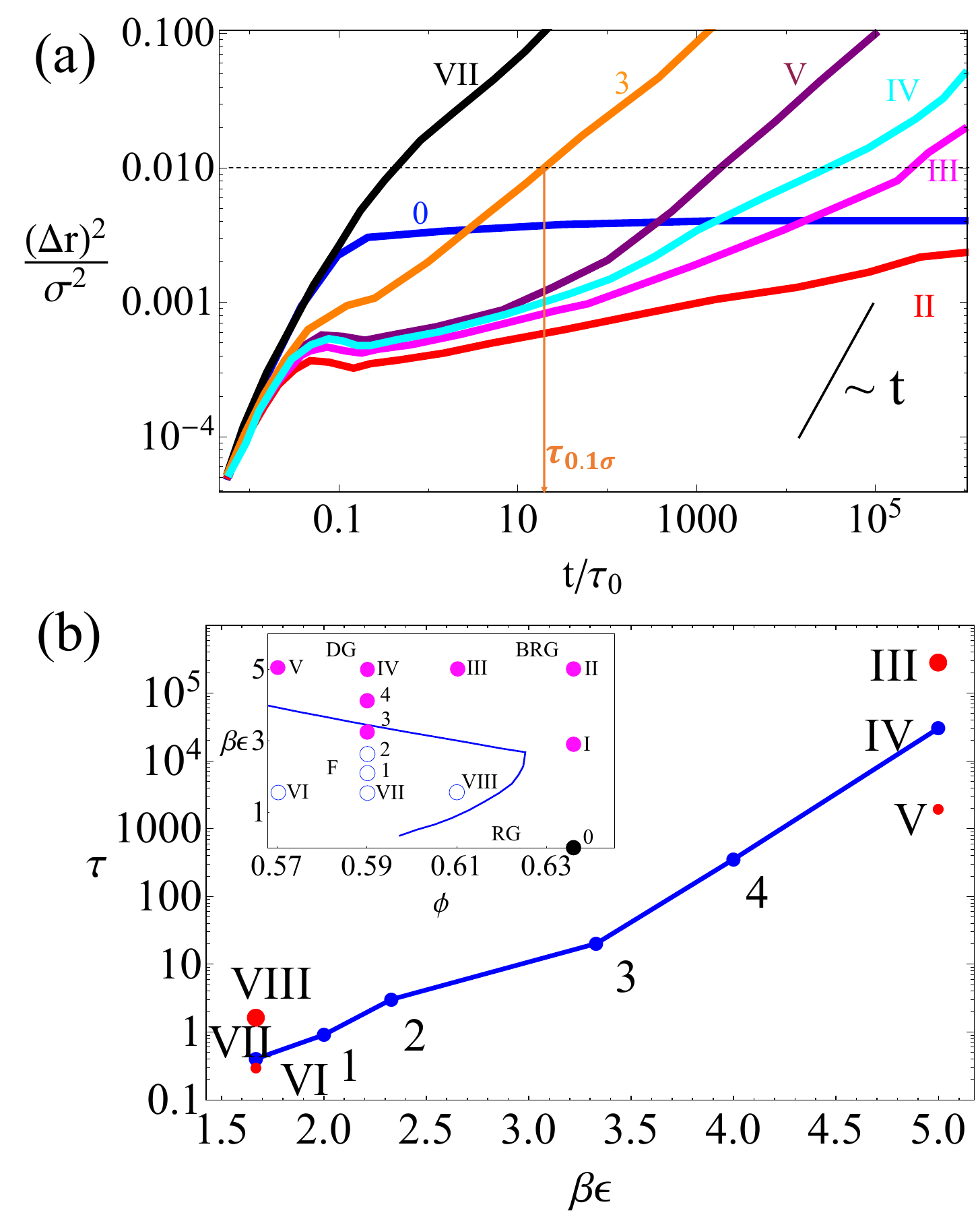}
    \caption{(a) Simulation MSD\cite{ZP2009} for selected state points as labeled in the inset of (b). Panel (b) is time required for MSD to reach $0.01\sigma^2$ (horizontal black dashed line in (a), illustrated for state 3 by the orange vertical arrow) plotted as a function of dimensionless attraction strength. Simulations of the filled (open) state points in the inset do not (do) reach the long time Fickian behavior.}
    \label{fig1}
\end{figure}
The simulation study also computed specific time correlation functions\cite{ZP2009}. The ``bonding correlation” quantifies the number of particles within a bonding distance $\sigma+a$ of a tagged particle, where “a” is the attraction range. The “caging correlation” is number of particles within a caging distance of $\sigma+R^*$ at time t averaged over all the particles, and $R^*$ is the MSD inflection point of the corresponding pure HS fluid and hence a fixed length scale criterion. By studying the time evolution of these and other quantities, an overarching conclusion was the bond-breaking time is always (much) shorter than the de-caging time defined from the MSD inflection point, and thus at longer times attractive glasses (BRG) are still dominated by steric caging. With increasing packing fraction and/or attraction strength, the difference between these two timescales grow, a form of “decoupling”. Unfortunately, the simulated time correlation functions do not decay enough to allow de-caging times to be extracted, and there was sufficient data for extracting only a few bond lifetimes. For the latter, the time scale modestly (of order a decade) increases with packing fraction at high attraction strength  (DG$\rightarrow$BRG) and at very high density as attraction strength grows (RG$\rightarrow$BRG). 
\par
Given the above, and our goal to objectively extract a characteristic time scale for “in cage” motion over a wide parameter space, we define a short timescale from the MSD data as  $\langle r^2(t=\tau_{0.1\sigma})\rangle=0.01\sigma^2$ (black dashed line in Fig 1a), the average time for a particle to move a relatively short distance ($0.1\sigma$). This displacement is well beyond a vibrational or bonding distance scale, and thus reflects bond breaking and the early stage of cage escape or “de-caging”. 
\par
From the simulation MSD data \cite{ZP2009} we determined  $\tau_{0.1\sigma}$ associated with 5 “trajectories” in the dynamic map of Fig (1b). For the vertical trajectories VI$\rightarrow$V, VII$\rightarrow$IV, VIII$\rightarrow$III, the time scales massively varies with increasing attraction by factors of $\sim$6700, 75,000, 200,000, respectively, with larger values for higher packing fractions. In strong contrast, the horizontal trajectories V$\rightarrow$III and VI$\rightarrow$VIII vary by factors of only $\sim$5 and 150, respectively. The timescales are plotted in Figure 1b as a function of strength of attraction. The vertical trajectory data can be fit to an Arrhenius form in attraction energy: $\tau_{0.1\sigma}(\beta\epsilon)=\tau_{0.1\sigma}(\beta\epsilon=1.67)\times e^{\alpha(\phi)(\beta\epsilon-1.67)}$. The effective activation energy is packing fraction dependent via the factor $\alpha(\phi)$ which we find is: $\alpha(VI\rightarrow V)\approx 2.64$, $\alpha(VII\rightarrow IV)\approx 3.37$, $\alpha(VIII\rightarrow III)\approx 3.62$. Thus, effective barriers are well beyond a single bond energy in magnitude and increase for denser fluids. These trends indicate the importance of many body effects and coupling of bond breaking and caging constraints. Figure 1 and the Table establish that changing attraction strength at fixed packing fraction produces much stronger changes of dynamics than vice-versa. 
\par
The simulation findings can potentially provide incisive tests of three fundamental aspects of the microscopic ECNLE-PDT approach. (i) Does the dynamic free energy faithfully capture the bond breaking and de-caging timescales throughout the wide parameter space? (ii) Does the qualitative difference for the structural relaxation time (and its enormous superiority) predicted based on the hybrid PDT construction of an effective force compared to its projection approximation analog \cite{agkss2019} persist for “in cage” motion? (iii) Is collective elasticity physics unimportant for smaller length-scale motion? Below we argue the answers to all these questions is yes.
\par
We consider a monodisperse sticky hard-sphere fluid of packing fraction $\phi=(\pi/6)\rho\sigma^3$,where $\rho$ is number density and $\sigma$ the  diameter with an exponential attractive pair potential: $\beta v(r)=-\beta\epsilon e^{-(r-\sigma)/a}; r\ge \sigma$, where $\beta\epsilon>0$. Equilibrium pair correlation functions are computed using Ornstein-Zernike theory with the Percus-Yevick closure \cite{hansen_mcdonald_2014}.
\begin{figure}
    \centering
    \includegraphics[scale=0.26]{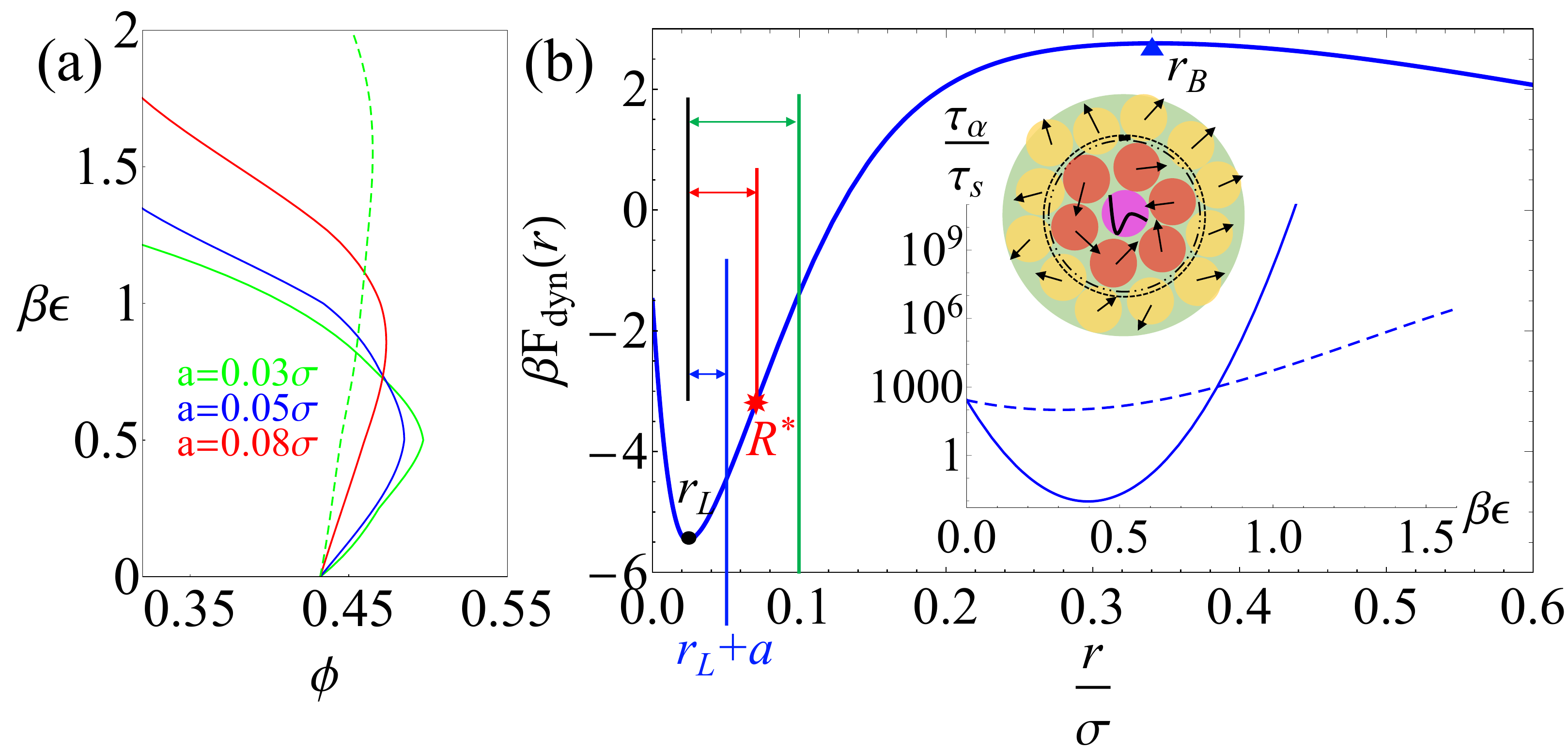}
    \caption{(a) Ideal (naïve MCT) kinetic arrest map based on the hybrid PDT effective force approximation\cite{agkss2019} for three short attraction ranges; one analogous result is shown using the standard projected effective force is shown for $a=0.03\sigma$ (green dashed curve). (b) Representative dynamic free energy as a function of particle displacement for the $\phi=0.58$ hard-sphere system. Important lengths indicated are: localization length ($r_L$), bond-breaking distance ($r_L+a$), location of maximum restoring ($R^*$), and barrier location ($r_B$). Inset shows the mean alpha relaxation time for $\phi=0.58$ and $a=0.03\sigma$ as a function of attraction strength for the projected effective force (dashed) and hybrid-PDT effective force (solid) approximations. Schematic of the conceptual basis of ECNLE theory involving small scale in-cage motion associated with longer range elastic fluctuations outside the cage.}
    \label{fig2}
\end{figure}
\par 
Trajectories are quantified by the angularly averaged scalar displacement of a tagged spherical particle from its t=0 initial position which obeys a NLE \cite{kss2005} $\zeta_s\frac{dr(t)}{dt}-\frac{\partial F_{dyn}(r(t))}{\partial r(t)}+\xi(t)=0$. Here, $\xi(t)$ is a white noise random force corresponding to the short time friction constant, and the gradient of the dynamic free energy $F_{dyn}(r)$ describes an effective force on a tagged particle due to the surroundings \cite{kss2005},
\begin{equation}
\begin{split}
    \beta F_{dyn}(r)=-3\ln(r/\sigma)-\rho\int\frac{d\vec{k}}{(2\pi)^3}\frac{|\vec{M}(k)|^2S(k)}{k^2(1+S^{-1}(k))}\\e^{-\frac{k^2r^2}{6}(1+S^{-1}(k))}
\end{split}
\end{equation}
where $\beta=(k_BT)^{-1}$ is inverse thermal energy, $S(k)=(1-\rho C(k))^{-1}$ is the static structure factor, $C(k)$ the direct correlation function, and $M(k)$ the Fourier transform of the effective interparticle force. In the standard projection approach \cite{Saltzman2003,kss2005} $|\vec{M}_{MCT}(k)|=kC(k)$, while for the new hybrid-PDT approach \cite{agkss2019},
\begin{equation}
    \vec{M}_{Hybrid}(k)=kC_0(k)\hat{k}+\int d\vec{r} g(r)\vec{f}_{att}(r)e^{-i\vec{k}.\vec{r}}
\end{equation}
where the subscript 0 indicates the pure hard sphere fluid quantity. 
\par
In general, beyond the  naïve (single particle) MCT transition the dynamic free energy has a (transient) localization length at $r=r_L$, and a barrier of height $F_B$ at $r=r_B$ (see Fig.2b). ECNLE theory includes longer range collective elastic fluctuations as a critical component of the coupled local-nonlocal alpha relaxation event \cite{mirigian2013,Steve12014}. Prior applications have focused on the long time structural relaxation process corresponding to the relatively large “jump distance” required to reach the dynamic free energy barrier. However, the physical ideas are general, and apply for any smaller length scale “uphill” motion from $r=r_L$ to $r=x$ corresponding to the displacement $\Delta r_x=x-r_L$. The collective elastic displacement field, $u(r)$, and its amplitude, follow as before \cite{mirigian2013,Steve12014}: $u(r,x)=\Delta r_{eff}(x)\Big(\frac{r_{cage}}{r}\Big)^2$, where $\Delta r_{eff}(x)=\frac{3}{32}\frac{(x-r_L)^2}{r_{cage}}$, for $r\ge r_{cage}\approx 1.3-1.5\sigma$. The corresponding elastic energy cost is \cite{mirigian2013}: $\beta F_{el}(x)\approx 12\phi K_0\Delta r_{eff}(x)^2(r_{cage}/\sigma)^3$, where $K_0=3k_BT/r_L^2$ is the harmonic spring constant of the dynamic free energy at its minimum. \par
The mean time required for particles to displace from $r=r_L$ to $r=x$ is calculated from Kramers theory \cite{kss2005},
\begin{equation}
    \frac{\tau_x}{\tau_s}=\sigma^{-2}e^{\beta F_{el}(x)}\int_{r_L}^x e^{\beta F_{dyn}(z)}dz\int_{r_L}^z e^{-\beta F_{dyn}(y)}dy
\end{equation}
where $\tau_s=\beta\zeta_s\sigma^2$ is the known \cite{mirigian2013} short time scale, here taken as the elementary time unit. We note that the hybrid PDT-ECNLE theory has been very recently employed \cite{agkss2019} to study aspects of the non-Gaussian parameter for dense attractive particle fluids including its maximum amplitude and corresponding time ($t^*$) which occur at a mean particle displacement ($R^*$) associated with the point of maximum force on the dynamic free energy \cite{Saltzman2008,saltzman2008anomalous} (Fig.2b). Predictions \cite{agkss2019} are consistent with simulation \cite{Reichaman2005}. Importantly, since $R^*$ is relatively small the collective elastic contribution was found to be negligible. Our present interests are on scales  $\sim R^*$ and shorter, and we again find elastic effects are unimportant. Hence, all results below reflect the core dynamic free energy physics plus the hybrid PDT effective force idea to treat short range attractive forces. 
\par
As relevant background, Fig. 2a shows naïve MCT localization boundaries based on the projection and hybrid PDT effective force approximations. They are qualitatively similar, but treating attractive forces explicitly results in a stronger re-entrant nose feature which intensifies for shorter range attractions since $f_{att}\propto\epsilon/a$ \cite{agkss2019}.  Figure 2b shows a sample dynamic free energy well beyond the naïve MCT crossover for the hard sphere fluid at $\phi=0.58$ The hybrid-PDT approach correctly\cite{agkss2019} accounts for its strong non-monotonic variation, including an almost 4 decade faster relaxation for the most glass-melted state which occurs at an attraction strength close to the nose of the ideal arrest boundary in Fig.2a. In qualitative contrast, only a hint of non-monotonicity is observed when the projected effective force is used, which disagrees with experiment and simulation\cite{agkss2019}.
\par
We now study the specific shorter length scale dynamical processes of present interest. (a) Bond-breaking timescale, $\tau_b$ , for displacing uphill on the dynamic free energy from $r=r_L$  to $x=r_L(\phi,\beta\epsilon,a/\sigma)+a$ (blue arrow in Fig. 2b). (b) De-caging timescale, $\tau_d$, where $x=R^*(\phi,\beta\epsilon,a/\sigma)$ with $R^*$ the location of maximum restoring force which occurs\cite{agkss2019,Saltzman2008,saltzman2008anomalous} at the displacement where there is an inflection point of the MSD(t)  (red arrow in Fig.2b). (c) Timescale to move a fixed distance $x=0.10\sigma$ defined as $\tau_{0.1\sigma}$ (green arrow in Fig.2b) per Fig. 1. The displacement associated with (a) is the smallest, while that of (b) and (c) are comparable with their relative magnitude depending on system state point. Since the dynamic free energy minimum at $r_L$ defines the displacement of the transiently localized bonded and caged state, it is a natural choice for the lower limit in Eq.(3) to address processes (a) and (b). For process (c), we have verified our calculations of $\tau_{0.1\sigma}$ are insensitive to further reduction of the lower limit, and the reason is that “downhill” motion from $r=0\rightarrow r_L$ is extremely fast.
\par
Our interests go beyond the simulation study \cite{ZP2009}, but to connect with it we note their MSD’s were largely determined at $\beta\epsilon=$0,1.67,5 with latter differing by factor of 3, and the middle value expected to be close to the nose feature per Fig. 1b.  This motivates our choice of parameters: an intermediate $\beta\epsilon=0.5$ close to the theory nose in Fig.2a, and a value 3 times larger, $\beta\epsilon=1.5$. The baseline attraction range and packing fraction are $3\%$ and 0.58, respectively, with a few results shown or discussed for other choices.
\begin{figure}
    \centering
    \includegraphics[scale=0.25]{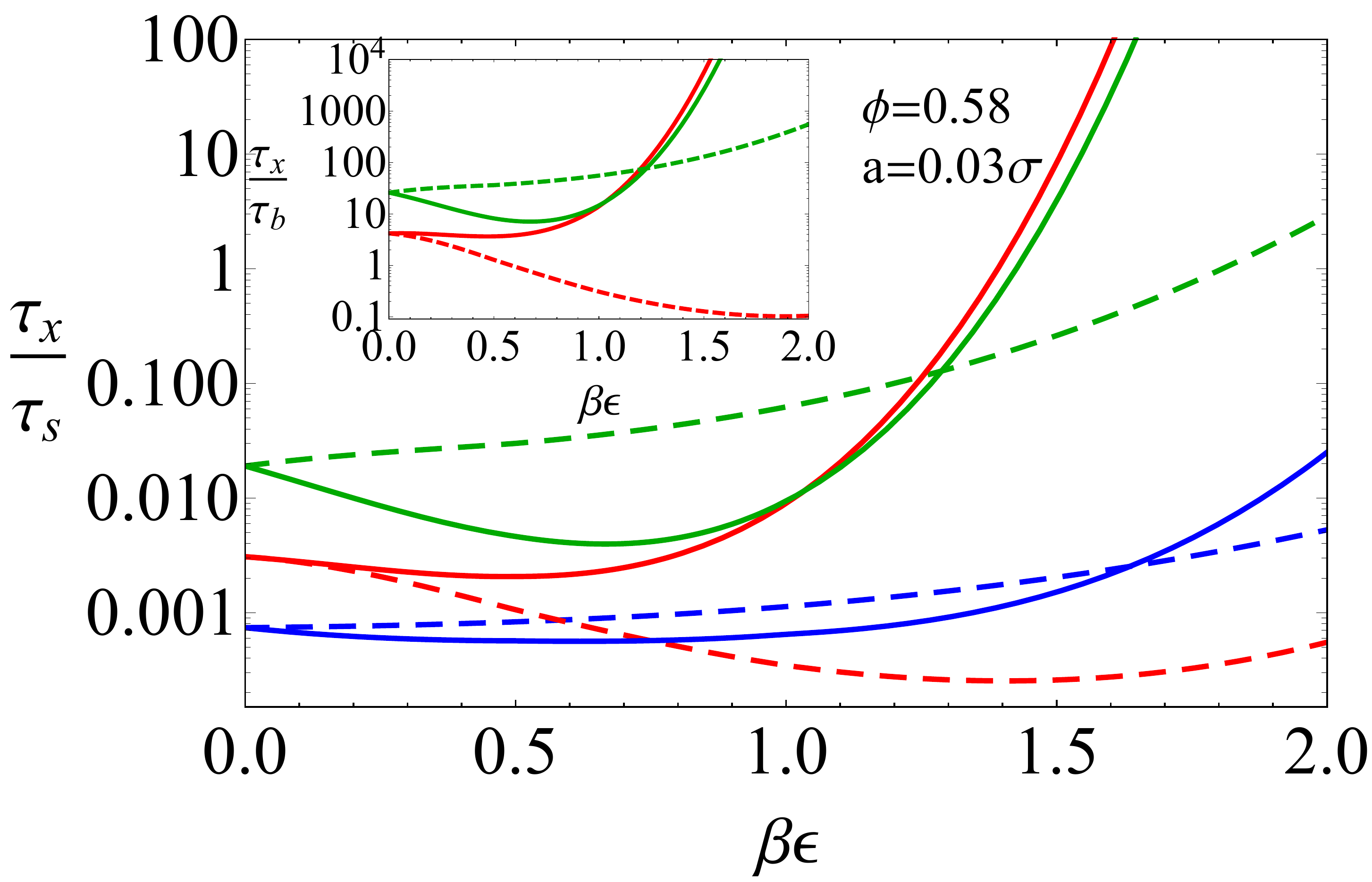}
    \caption{Non-dimensionalized timescales of bond-breaking (blue), de-caging (red) and fixed distance displacement of $0.10\sigma$ (green) for  $\phi=0.58$ and attraction range  $a=0.03\sigma$. Inset: ratio of de-caging (red) and fixed displacement time scales to the bond-breaking timescale for the same packing fraction and attraction range. Solid and dashed curves indicate the hybrid-PDT and projected effective force based ECNLE theory results, respectively.}
    \label{fig3}
\end{figure}
\par
The main frame of Figure 3 presents the three time scales in a common elementary unit as a function of attraction strength corresponding to vertical trajectories in Fig.1 that transverses the RG, F, and BRG states. Results based on both the hybrid PDT and projected effective force approximations are shown. The inset shows “decoupling ratios”, defined as the ratio of the two slower time scales to the bond lifetime. There are multiple notable trends.
\par
For the hybrid-PDT based theory all timescales show weakly non-monotonic behavior, enormously gentler than predicted \cite{agkss2019} for the alpha time. For attraction strength beyond the ‘nose’ in Fig.2a, the de-caging time and $\tau_{0.1\sigma}$ are nearly the same since the corresponding displacements are close. The bond-breaking time varies much more weakly with attraction strength given the small displacement involved. Based on additional calculations, we find that the qualitative nature of the weak non-monotonic variation of timescales, weaker dependence of bond-breaking time compared to the other two timescales, and quantitative similarity of $\tau_{d}$ and $\tau_{0.1\sigma}$ beyond the ‘nose’, are all independent of the precise value of (high) packing fraction and (short) attraction range. The weak non-monotonicity in the main frame of Fig.3 can be tested in future simulations that employ a finer grain variation of attraction strength.
\par
The inset of Fig. 3 shows the decoupling ratios computed based on the hybrid PDT approach grow with attraction strength, and also very strongly (not shown) with packing fraction. For example, at $\beta\epsilon=1.5$ the decoupling ratio is $\sim 10^2, 10^4, 10^6$ for packing fractions (spanning the DG to BRG regimes) of 0.56, 0.58, 0.60, respectively. This large growth with modest packing fraction increase seems consistent with the evolution from a DG to BRG \cite{ZP2009}. The results based on the projected effective force are completely different, exhibiting little change or even opposite trends with attraction strength, and do not seem physical.  
\par
\begin{table*}
Table: Comparison of simulation results\cite{ZP2009} for the overall range of variation of $\tau_{0.1\sigma}$ along different trajectories in Figure 1 with the corresponding hybrid PDT ECNLE theory predictions for a monodisperse sphere model with an exponential attraction of 3\% range. \\
\begin{center}
\begin{tabular}{|p{4.2cm}|p{4.2cm}|p{4.2cm}|p{4.2cm}|}\hline
\multicolumn{2}{|p{5cm}|}{\textbf{Trajectory}} & \multicolumn{2}{|p{5cm}|}{\textbf{Ratio of Timescales $\tau_{0.1\sigma}$}} \\ \hline
Simulation\footnote{State in Fig. (1b) inset} & Theory\footnote{State=$(\phi,\beta\epsilon)$} & Simulation & Theory \\ \hline
VI $\rightarrow$V & (0.56,0.55) $\rightarrow$ (0.56,1.65)& $6.7\times 10^3$ & $4.4\times 10^3$ \\ \hline
VII$\rightarrow$ IV & (0.58,0.55) $\rightarrow$ (0.58,1.65)& $7.5\times 10^4$ & $2.6\times 10^4$ \\ \hline
VIII$\rightarrow$ III & (0.60,0.55) $\rightarrow$ (0.60,1.65)& $2\times 10^5$ & $5.4\times 10^5$ \\ \hline
VI$\rightarrow$ VIII & (0.56,0.55) $\rightarrow$ (0.60,0.55)& $5\times 10^0$ & $5.1\times 10^0$ \\ \hline
V$\rightarrow$ III & (0.56,1.65) $\rightarrow$ (0.60,1.65)& $1.5\times 10^2$ & $6.0\times 10^2$ \\ \hline 
\end{tabular}
\end{center}
\end{table*}
\par
The simulation\cite{ZP2009} vertical trajectory results in the Table can be compared to our calculations. States VI, VII and VIII are close to the ‘nose’ in the kinetic arrest diagram. States, V, IV and III are reached from VI, VII and VIII by increasing the attraction strength by a factor of ~3 at $\phi=$0.57, 0.59 and 0.612, respectively. We map\footnote{Particle size polydispersity speeds up relaxation. Since we employ a monodisperse model, the precise mapping of our packing fraction to that of the simulation has uncertainty. We crudely account for this by reducing the simulation packing fraction by 0.01 in our calculations. Though not rigorous, we believe this is a reasonable attempt to account for this feature} $\phi=0.57$ of the simulation mixture model to 0.56 in our monodisperse model; the other two states then roughly correspond to $\phi=0.58$ and 0.60, respectively. The nose in the kinetic arrest diagram for $a=0.03\sigma$ is at $\beta\epsilon\sim 0.55$, thus a factor of $\sim 3$ increase in attraction strength is achieved by increasing $\beta\epsilon=0.55\rightarrow 1.65$. These “calibrated” parameter choices allow us to plausibly compare the hybrid-PDT based calculations and simulations values for  $\tau_{0.1\sigma}$. The results are given in the Table. We uniformly find consistent, and in some cases nearly quantitative, agreement between theory and simulation. This seems especially significant since the overall time scale change can be as large as 5 decades, and the degree of variation is highly trajectory specific.
\begin{figure}[h]
    \centering
    \includegraphics[scale=0.25]{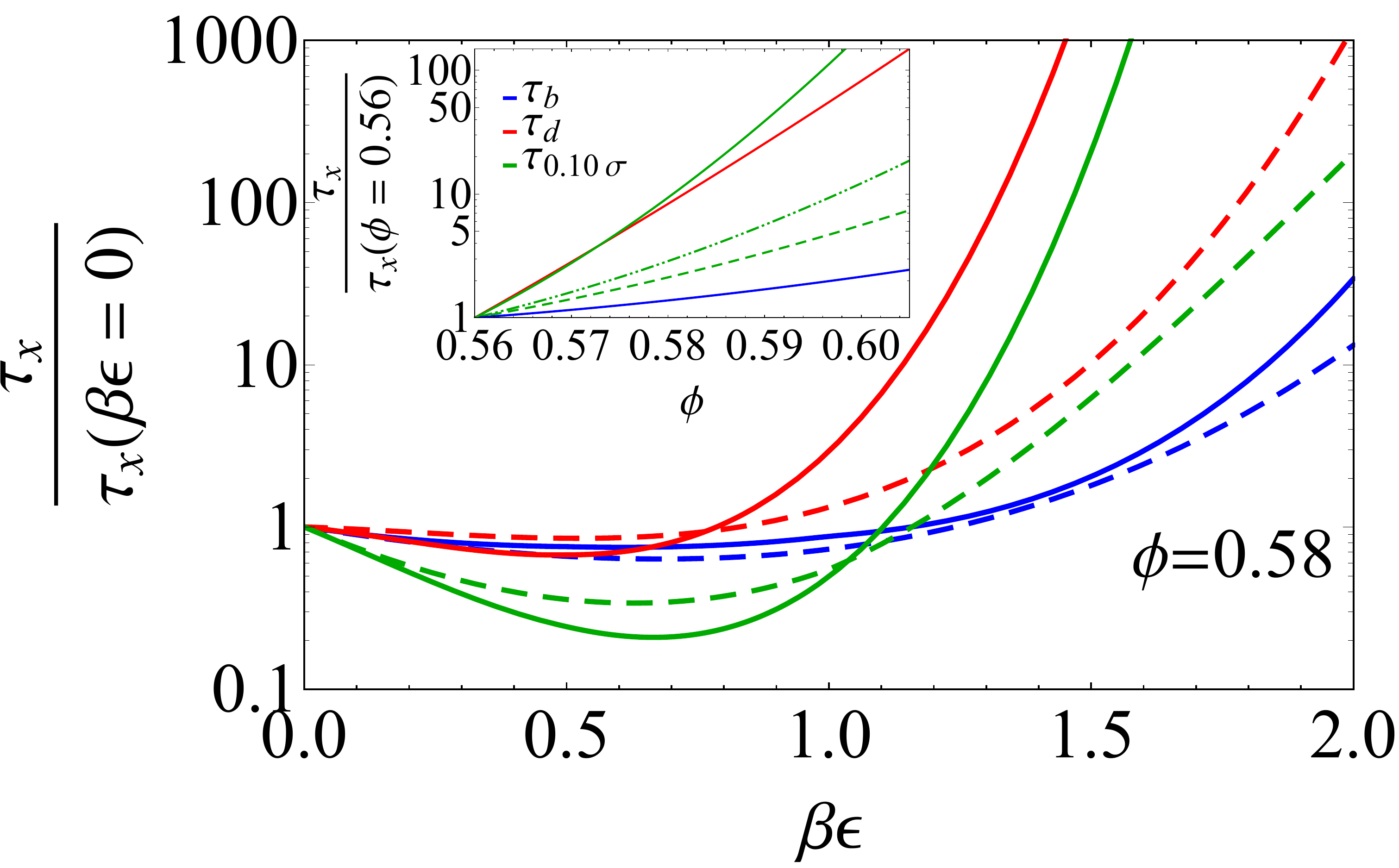}
    \caption{Same 3 hybrid-PDT based timescales as in Fig.3 normalized by their corresponding pure hard sphere fluid values as a function of attraction strength for ranges of $a=0.03\sigma$ (solid) and $a=0.08\sigma$ (dashed). Inset: timescales normalized by their value at a packing fraction of 0.56 for a range of $a=0.03$ as a function of higher volume fractions for:  (i) the fixed displacement timescale (green curves) at attraction strengths of 0 (hard sphere, dot-dashed), 0.50 (~around nose, dashed) and 1.50 (well above the nose, solid), and (ii) the other 2 timescales just at an attraction strength of 1.50}
    \label{fig4}
\end{figure}
The main frame of Fig.4 presents hybrid PDT based results which establish how the three timescales grow relative to their hard sphere values as a function of attraction strength for two ranges of 3\% and 8\%. The dependence is slightly non-monotonic which weakens as the attractive force range becomes longer. \par
The inset of Figure 4 shows results for a horizontal trajectory in Fig.1 that quantifies how the 3 timescales (normalized to their values at $\phi=$0.56) evolve at various fixed attraction strengths over the “wide” range of $\phi\in \{0.56, 0.61\}$. For context, we recall that for pure hard spheres as $\phi=0.56\rightarrow0.61$ ECNLE theory predicts \cite{Steve12014} the structural relaxation time grows by $\sim9$ decades. The three reduced timescales are plotted for different attraction strengths of 0, 0.50 (intermediate, near nose $\beta\epsilon$ in Fig.(2a)), and 1.5 (high $\beta\epsilon$, spans the BRG to DG region). Bond-breaking times exhibit a very weak packing fraction dependence compared to the other two timescales, in qualitative accord with the simulation \cite{ZP2009}. For high $\beta\epsilon=1.5$, bond breaking times increase only by a small factor of $\sim3$, whereas the other two timescales grow by $\sim$2 decades. The relative growth of $\tau_d$ and $\tau_{0.1\sigma}$ are strongly dependent on attraction strength. All timescales increase in a nearly exponential manner with packing fraction. Importantly, the Table shows the predicted packing fraction dependences of the fixed displacement timescale are in good accord with simulation results at intermediate and strong attractions strengths.
\par
We have extended the microscopic ECNLE plus hybrid PDT theory to quantitatively study attractive particle dynamics in ultra-dense fluids on smaller “in-cage” length and timescales where collective elasticity effects are unimportant. Our results for the de-caging, bond breaking, and fixed displacement timescales are in good accord with simulation\cite{ZP2009} over a wide range of attraction strengths and packing fractions. This supports the underlying spatially-resolved dynamic free energy concept in the theory. Testable predictions are made for attraction ranges and packing fractions not yet simulated. Orders of magnitude decoupling of the bond breaking time from the de-caging time is predicted, which is greatly enhanced with increasing density. The present work also establishes that our new treatment of attractive forces is critical to understand the “uphill” (thermally activated) shorter time and length scale motions, as recently demonstrated for the long time barrier hopping process\cite{agkss2019}. Future directions include applying the theory to other physical bond forming soft matter systems, stochastic trajectory solution of NLE theory to predict time correlation functions \cite{Saltzman2008}, and incorporation of deformation to study the “double yielding” rheological phenomenon\cite{Pham104,Koumakis,Pham2008} in ultra-dense attractive colloidal suspensions. 
\begin{acknowledgments}
This work was supported by DOE-BES under Grant DE-FG02-07ER46471 administered through the Materials Research Laboratory at UIUC. 
\end{acknowledgments}
\bibliographystyle{apsrev4-2}
\bibliography{apssamp.bib}

\end{document}